\documentclass{birkmult}
 \theoremstyle{definition}
 
 \theoremstyle{remark}

 \numberwithin{equation}{section}

\begin{document}
%
%
%
%
%
%
%
%
%
\title[Time paradox in Quantum Gravity]
 {Time paradox in Quantum Gravity}
\author[Alfredo Mac\'{\i}as and Hernando Quevedo]{Alfredo Mac\'{\i}as}

\address{Departamento de F\'{\i}sica,\\ Universidad Autonoma Metropolitana--Iztapalapa,\\
A.P. 55--534, M\'exico D.F. 09340, M\'exico}

\email{amac@xanum.uam.mx}

\author{Hernando Quevedo}
\address{Instituto de Ciencias Nucleares,
Universidad Nacional Aut\'onoma de M\'exico\\
A.P. 70-543,  M\'exico D.F. 04510, M\'exico}
\email{quevedo@nuclecu.unam.mx}



\begin{abstract}
The aim of this work is to review the concepts of time in quantum
mechanics and general relativity to show their incompatibility. We
show that the absolute character of Newtonian time is present in
quantum mechanics and also partially in quantum field theories
which consider the Minkowski metric as the background spacetime.
We discuss the problems which this non-dynamical concept of time
causes in general relativity that is characterized by a dynamical
spacetime.
\end{abstract}

\maketitle
\section{Introduction}

Our present description of the laws of physics may be
characterized as obtained from two types of constituents. The
first type of constituent are theoretical frameworks which  apply
to {\em all} physical phenomena at {\em any} instant. These
``universal''  or ``frame''  theories are Quantum Theory, i.e.,
all matter is of microscopic origin, Special and General
Relativity, i.e., all kinds of matter locally have to obey the
principles of Lorentz symmetry and behave in the same way in
gravitational fields, and Statistical Mechanics which is a method
to deal with all kinds of systems for a large number of degrees of
freedom. The second type of constituent is non--universal and
pertains to the description of the four presently--known
interactions: the electromagnetic, the weak, the strong, and the
gravitational. The first three interactions are all described
within a single formalism, in terms of a gauge theory. So far only
gravity has not been successfully included into that scheme. One
reason for that might be that gravity appears on both sides: it is
an interaction but it is at the same time also a universal theory.
Universal theories like relativity and gravity are geometric in
origin and do not rely on the particular physical system under
consideration, whereas a description in terms of a particular
interaction heavily makes use of the particular particle content.
Therefore, gravity plays a distinguished role which may be the
reason for the difficulty encountered in attempting to unify the
other interactions with gravity and attempting to quantize gravity
\cite{aclmm05,kiefer}.

The concepts of time in quantum mechanics and general relativity
are drastically different from each other.

One one hand, time in quantum mechanics is a Newtonian time, i.e.,
an absolute time. In fact, the two main methods of quantization,
namely, canonical quantization method due to Dirac and Feynman's
path integral method are based on classical constraints which
become operators annihilating the physical states, and on the sum
over all possible classical trajectories, respectively. Therefore,
both quantization methods rely on the Newton global and absolute
time. The absolute character of time in quantum mechanics results
crucial for its interpretation, i.e., matrix elements are
evaluated at fixed time, and the internal product is unitary,
i.e., conserved in time, and it implies conservation of the total
probability. Therefore, time is part of the classical background,
which is needed for the interpretation of measurements. Moreover,
the introduction of a time operator in quantum mechanics is thus
problematic. The time parameter appears explicitly in the
Schr\"odinger equation, together with the imaginary unit. Since
time is absolute can be factorized, for instance, in the canonical
quantization, reducing the quantization problem to the
construction of a Hilbert space for stationary states.

The transition to (special) relativistic quantum field theories
can be realized by replacing the unique absolute Newtonian time by
a set of timelike parameters associated to the naturally
distinguished family of relativistic inertial frames. In this
manner, the time continues to be treated as a background
parameter.

On the other hand, time in general relativity is dynamical and
local. Hence, it is not an absolute time. The geometry of
spacetime influences material clocks in order to allow them to
display proper time, and the clocks react on the metric changing
the geometry. Therefore, the metric itself results to be a clock,
and the quantization of the metric can be understood as a
quantization of time \cite{zeh01}.

The above mentioned quantization methods, when applied to general
relativity lead to the Wheeler--DeWitt equation
\cite{halliwell88}. It is well known that, as a second order
functional differential equation, the Wheeler--DeWitt equation
presents familiar problems when one tries to turn the space of its
solutions into a Hilbert space \cite{Kuchar}.

In full, general relativity does not seem to possess a natural
time variable, while quantum theory relies quite heavily on a
preferred time. Since the nature of time in quantum gravity is not
yet clear, the classical constraints of general relativity do not
contain any time parameter, and one speaks of the {\em time
paradox}.

The aim of the present work is to review the concepts of time in
both quantum mechanics and general relativity. Our understanding
of time is in the context of the canonical quantization approach
to quantum mechanics and quantum field theory. This is why we
review in section \ref{sec:canqua}  the axioms of canonical
quantization, emphasizing the role of time at each step. Then, in
section \ref{sec:genrel}, we discuss the role of the time
parameter in general relativity and establish its dynamical
character. Sections \ref{sec:mini} and \ref{sec:midi} are devoted
to brief descriptions of how time enters the problem of canonical
quantization on minisuperspaces and midisuperspaces, respectively.
Section \ref{sec:probtime} contains  a discussion on the main
approaches used to attack the problem of time. Finally, section
\ref{sec:con} is devoted to the conclusions.

\section{Time in canonical quantization}
\label{sec:canqua} Quantum theory is based on a certain procedure
of quantization of a classical system which consists of a series
of axioms. The standard and most used procedure is canonical
quantization, whose starting point is the Hamiltonian describing
the classical system. It is interesting  that, like any other
physical theory, there is no proof for quantum theory. The only
thing we know for sure is that the experimental observations of
Nature do not contradict the predictions of quantum theory, at
least within the range of measurements accessible to current
experimental devices. In canonical quantization time plays a very
important role in all the axioms which are postulated as the
fundamentals of this method. First, the mere fact that one needs
to know the Hamiltonian of the system implies that a certain time
parameter has to be chosen in order to define the variables in
phase space. To be more specific let us briefly recall the main
axioms of canonical quantization.

In the case of quantum mechanics for a system with only {\em
bosonic degrees of freedom} these axioms can be stated as follows:

\noindent I) There exists a Hilbert space {\bf H} for the quantum
system and the elements of {\bf H} are the quantum states
$|\psi\rangle$ of the system. The Hilbert space is supposed to be
equipped with an inner product, i.e. a positive definite Hermitian
norm on {\bf H}. Often the inner product of two elements
$|\phi\rangle$ and $|\psi\rangle$ of {\bf H} is denoted as
$\langle\phi|\psi\rangle$.

\noindent II) A classical observable $A$ is replaced by a
Hermitian operator ${\hat A}$ acting on elements of {\bf H}. When
the observable $A$ is measured, the result must coincide with one
of the eigenvalues of ${\hat A}$. It is also assumed that for any
physical state $|\psi\rangle \in$ {\bf H}, there exists an
operator for which the state $|\psi\rangle$ is one of its
eigenstates.

\noindent III) If $q_i$ and $p_j$ ($i,j=1,2, ... n = $ number of
bosonic degrees of freedom of the system), are the variables in
phase space ${\bf R}^{2n}$, the corresponding operators must obey
the commutation relations at a fixed time $t$
\begin{equation}
[\hat q_i, \hat q_j] = 0, \quad [\hat p_i, \hat p_j] = 0, \quad
[\hat q_i, \hat p_j] = i \delta_{ij}\ ,
\end{equation}
where we are using units with $\hbar = 1$.  This axiom can be
generalized to include the case of phase spaces other than  ${\bf
R}^{2n}$ (see, for instance, \cite{wood,isham,ash}).

\noindent IV) If $\hat A$ does not depend explicitly on time, its
evolution in time is determined by Heisenberg's evolution
equation:
\begin{equation}
\frac{d \hat A}{d t} = \frac{1}{i}[\hat A , \hat H]\ .
\end{equation}
The formal solution of this equation $\hat A (t) = e^{i \hat H t}
\hat A (0) e^{-i \hat H t}$ can be used to obtain the equivalent
Schr\"odinger picture in which the operators $\hat A (0)$ are
time--independent and instead the states become time--dependent
through the unitary transformation $|\psi(t)\rangle = e^{- i \hat
H t} |\psi\rangle$. Then, the evolution of a state of the physical
system turns out to be determined by the Schr\"odinger equation
\begin{equation}
i \frac{d}{d t} |\psi(t)\rangle = \hat H |\psi(t)\rangle\ .
\label{hei}
\end{equation}

\noindent V) In general, the observation of $A$ in a physical
system at a fixed time $t$ yields random results whose expectation
value is given by
\begin{equation}
\langle A\rangle_t = \frac{\langle\psi|\hat A
(t)|\psi\rangle}{\langle\psi|\psi\rangle} \ . \label{sch}
 \end{equation}

These are the axioms that lie on the basis of canonical
quantization for classical systems with a finite number of degrees
of freedom. The time parameter $t$ plays a very important role in
determining the phase space, i.e. the choice of canonical
positions $q_i$ and momenta $p_i$.  Fortunately, the time used in
classical mechanics is the absolute Newtonian time which is
defined up to constant linear transformations. Thus, the conjugate
momenta are determined up to a multiplicative constant which does
not affect the main structure of the phase space. This absolute
time is then used with no changes in the quantization scheme
described in the above axioms.

Time enters explicitly in axioms III and V, since the commutation
relations must be satisfied at a given moment in time and the
results of any observation lead to expectation values which are
well--defined only if time is fixed. This crucial role of time can
be rephrased in terms of the wave function. Indeed, if we define
the wave function $\psi(t,x)$ as $\psi(t,x)= \langle
x|\psi(t)\rangle$, fixing its normalization, means that it must be
normalized to one at a fixed time.

The equation of evolution (\ref{hei}) represents changing
relations amongst the fundamental entities (operators) of this
construction. This equation indicates which operator has to be
used to describe the physical system at a given time. When time
changes, Heisenberg's equation explains which operator in Hilbert
space corresponds to the new state of the physical system.

These observations indicate that in canonical quantization time is
an ``external" parameter. It is not a fundamental element of the
scheme, but it must be introduced from outside as an absolute
parameter which coincides with the Newtonian time. Since there is
no operator which could be associated with time, it is {\em not}
an observable.

The transition to quantum field theory is performed in a
straightforward manner, although many technical details have to be
taken into account \cite{waldbook}. The main variables are now the
value of the field $\varphi({\bf x})$ at each spatial point and
the conjugate momentum $\pi({\bf x})$ for that particular value.
The collection of all the values of the field, together with the
values of the conjugate momenta, represents the variables of the
new phase space. Axioms I -- V are then postulated for the
corresponding phase space variables. Some changes are necessary in
order to consider the new ``relativistic" time. In particular, the
commutation relation
\begin{equation}
[\hat \varphi(x), \hat \varphi(y)]= 0
\end{equation}
is valid for any spacetime points $x$ and $y$ which are spacelike
separated. The main difference in the treatment of fields is that
the time parameter is that of special relativity. Instead of the
absolute Newtonian time, we now have a different parameter
associated to each member of the distinguished class of inertial
frames. The two absolute concepts of Newtonian physics, i.e. space
and time, are now replaced by the single concept of spacetime.
Nevertheless, in special relativity spacetime retains much of the
Newtonian scheme. Although it is not possible to find an absolute
difference between space and time, spacetime is still an element
of the quantum theory which does not interact with the field under
consideration. That is to say, spacetime remains as a background
entity on which one describes the classical (relativistic) and
quantum behavior of the field.

In other words, one could say that an observer with the ability to
``see" only the physical characteristics of spacetime cannot
determine if he/she is ``living" on a spacetime with a classical
or a quantum field.  Spacetime in quantum field theory is
therefore an  external entity like the absolute external time in
quantum mechanics. The dynamics of the field does not affect the
properties of spacetime which is therefore a nondynamical element
of the theory.

\section{Time in general relativity}
\label{sec:genrel}

To implement the canonical quantization procedure in general
relativity one needs to find the classical Hamiltonian. As
mentioned above, such a formulation requires an explicit choice of
time or, equivalently, a slicing of spacetime into spatial
hypersurfaces associated to the {\em preferred} chosen time. This
is the Arnowitt--Deser--Misner (ADM) \cite{adm} approach which
splits spacetime into space and time. The pseudo--Riemannian
manifold describing the gravitational field is therefore
topologically equivalent to ${\bf R}\times \Sigma_t$, where ${\bf
R}$ represents the ``time axis", and $\Sigma_t$ are constant--time
hypersurfaces, each equipped with a set of three coordinates
$\{x^i\}$ and a non--degenerate 3--metric $q_{ij}$. The
relationship between the local geometry on $\Sigma_t$ and the
4--geometry can be recovered by choosing an arbitrary point on
$\Sigma_t$ with coordinates $x^i$ and displacing it by an
infinitesimal amount $dt$ normal to $\Sigma_t$. The result of this
infinitesimal displacement induces an infinitesimal change in
proper time $\tau$,  which can be written as $d\tau = N dt$, where
$N=N(x^\mu)$ is the lapse function, and an infinitesimal change in
spatial coordinates, which can be written as $x^i(t+dt) = x^i(t) -
N^i dt$, where $N^i = N^i(x^\mu)$ is the shift vector. Then the
4--dimensional interval connecting the starting $x^i$ and ending
$x^i + dx^i$ points of this infinitesimal displacement is given by
the ADM--metric
\begin{equation}
ds^2 = - N^2 dt^2 + q_{ij} (dx^i + N^i dt)( dx^j + N^j dt) \ .
\label{metadm}
\end{equation}
Notice that this splitting of spacetime explicitly depends on the
choice of the time parameter $t$. Indeed, the tensorial quantities
$N$, $N^i$, and $q_{ij}$ can be given different values by means of
a general diffeomorphism.

The Einstein--Hilbert action on a manifold $M$ with vanishing
cosmological constant reads
\begin{equation}
S_{EH} = \frac{1}{16\pi G} \int L d^4x = \frac{1}{16\pi G}\int_M
\sqrt{-g} R d^4x\ \pm \frac{1}{8\pi G} \int_{\partial M} \sqrt{q}
K d^3 x \label{ehaction} \, ,
\end{equation}
in terms of (\ref{metadm}) it becomes a function of the intrinsic
metric $q_{ij}$ and its derivatives of first order in time. The
boundary term in (\ref{ehaction}) is necessary in the variation to
cancel terms that arise after integrating by parts \cite{york}. It
is positive (negative) in case of spacelike (timelike)
 components of $\partial M$ and vanishes when the manifold is spatially compact.
The phase space is then constructed by means of the configuration
space variables $q_{ij}$ and their canonically conjugate momenta
$\pi^{ij} =
\partial L/\partial (\partial_t q_{ij})$ which are related
to the extrinsic curvature of the 3--dimensional hypersurface
$\Sigma_t$ as embedded in  the 4--dimensional spacetime. The
resulting Hamiltonian turns out to be that of a constrained
system, indicating that the phase space variables are not
independent. A straightforward calculation shows that the
Einstein--Hilbert action can be written as (dropping boundary
terms)
\begin{equation}
S_{EH}= \int d t \int_\Sigma d^3 x \left( \pi^{ij} \partial_t
q_{ij} - N H_\bot -N^i H_i \right) \ .
\end{equation}
Since this action does not contain time derivatives of $N$ and
$N^i$, their variation leads to the Hamiltonian constraint
(super--Hamiltonian constraint)
\begin{equation}
H_\bot := 16\pi G G_{ijkl} \pi^{ij} \pi^{kl}  - \frac{1}{16\pi G}
\sqrt{q}\ ^{(3)} R \ = 0, \label{ham}
\end{equation}
and the constraint of spatial diffeomorphisms (super--momentum
constraint)
\begin{equation}
H^i ({\bf x}) =
 - 2\ ^{(3)}\nabla_j \pi^{ij} = 0\ .
\label{3diff}
\end{equation}
Here $q$ is the determinant and $^{(3)}R$ the curvature scalar of
the 3--metric $q_{ij}$. The covariant derivative with respect to
$q_{ij}$ is denoted by $^{(3)}\nabla_j$. The DeWitt supermetric is
defined as
\begin{equation}
G_{ijkl} := \frac{1}{2\sqrt{q}}(q_{ik}q_{jl} + q_{jk} q_{il} -
q_{ij}q_{kl}) \ .
\end{equation}
Einstein's field equations are now the standard Hamilton equations
for the corresponding action with the Poisson brackets defined
according to
\begin{equation}
\{q_{ij}({\bf x}),\pi^{kl}({\bf x}^\prime)\} = \delta^k_{(i}
\delta^l_{j)} \delta({\bf x},{\bf x}^\prime) \ .
\end{equation}

This special slicing, in which the structure of the spatial
hypersurfaces $\Sigma_t$ is determined as the $t=$ const.
surfaces, leads to the first computational complication for the
algebra of diffeomorphisms. In fact, the diffeomorphism invariance
in the starting 4--dimensional spacetime is well defined in terms
of the corresponding Lie group. When this spacetime diffeomorphism
invariance is projected along and normal to the spacelike
hypersurfaces $\Sigma_t$, one obtains
\begin{equation}
\{H_i({\bf x}),H_j({\bf x}^\prime)\} = H_i({\bf x}^\prime)
\partial^{\bf x}_j \delta({\bf x}, {\bf x}^\prime) - H_j({\bf x})
\partial^{{\bf x}^\prime}_i \delta({\bf x}, {\bf x}^\prime) \ ,
\label{alg1}
\end{equation}
\begin{equation}
\{H_i({\bf x}),H_\bot({\bf x}^\prime)\} = H_\bot({\bf x})
\partial^{\bf x}_i \delta({\bf x}, {\bf x}^\prime) \ ,
\label{alg2}
\end{equation}
\begin{equation}
\{H_\bot({\bf x}),H_\bot({\bf x}^\prime)\} = q^{ij}({\bf x})
H_i({\bf x}) \partial^{{\bf x}^\prime}_j \delta({\bf x}, {\bf
x}^\prime) - q^{ij}({\bf x}^ \prime)H_i({\bf x}^\prime)
\partial^{{\bf x}}_j \delta({\bf x}, {\bf x}^\prime) \ .
\label{alg3}
\end{equation}

The fact that the right--hand side of Eq.(\ref{alg3}) contains the
3--metric explicitly implies that the projected algebra of
constraints is not a Lie algebra. This is a consequence of the
choice of time which leads to  considerable computational
complications for quantization \cite{dewitt}. One could try to
choose a specific gauge in accordance to the invariance associated
with the algebra (\ref{alg1})--(\ref{alg3}), then solve the
constraints (\ref{ham}) and (\ref{3diff}), and finally quantize
the resulting system with the ``true" degrees of freedom. It turns
out that in general the final equations are tractable only
perturbatively, and lead to ultraviolet divergences (for further
details see, for example, \cite{isham99,carlip01}).

An alternative approach consists in applying the canonical
quantization procedure to the complete collection of variables in
phase space. The variables $q_{ij}$ and $ \pi^{jk}$ are declared
as operators $\hat q _{ij}$ and $ \hat \pi ^{jk}$ which are
defined on the hypersurface $\Sigma_t$ and satisfy the commutation
relations
\begin{eqnarray}
& &[\hat q _{ij}({\bf x}), \hat q _{kl}({\bf x}^\prime)] =0 \ ,
\cr & &[\hat \pi ^{ij}({\bf x}), \hat \pi ^{kl}({\bf x}^\prime)]
=0 \ , \cr & &[\hat q _{ij}({\bf x}), \hat \pi ^{kl}({\bf
x}^\prime)] = i \delta^k_{(i}\delta^l_{j)}
    \delta({\bf x}, {\bf x}^\prime)\ .
    \label{comm}
\end{eqnarray}
According to Dirac's quantization approach for constrained
systems, the operator constraints must annihilate the physical
state vectors, i.e.,
\begin{equation}
\hat H _\bot \Psi[q] = 0 \ , \label{hamc}
\end{equation}
\begin{equation}
 \hat H _ i \Psi[q] = 0\, ,
\label{diffc}
\end{equation}
at all points in $\Sigma_t$. If the standard representation
 \begin{equation}
\hat q _{ij} \Psi[q] := q_{ij} \Psi[q]\, , \qquad \hat
\pi^{ij}\Psi[q] := -i \frac{\delta\Psi[q]}{\delta q_{ij}} \, ,
\end{equation}
is used, the constraint $\hat H _i \Psi[q] =0$ requires that
$\Psi[q]$ behaves as a constant under changes of the metric
$q_{ij}$ induced by infinitesimal diffeomorphisms of the
3-dimensional hypersurface $\Sigma_t$. In this specific
representation the Hamiltonian constraint (\ref{hamc}) becomes the
Wheeler--DeWitt equation
\begin{equation}
- 16\pi G \ G_{ijkl} \frac{\delta  ^2 \Psi[g]}{\delta q_{ij}
\delta q_{kl}} - \frac{1}{16\pi G}\ ^{(3)}R \Psi[g] = 0 \ .
\label{wdw}
\end{equation}
In canonical quantization this is considered as the main dynamical
equation of the theory, since classically the function(al)
$H_\bot$ is associated with the generator of displacements in
time--like directions. That is to say, $H_\bot$ is the generator
of the classical evolution in time. By analogy with quantum
mechanics or quantum field theory one expects that the
Wheeler--DeWitt equation (\ref{wdw}) determines the evolution
among quantum states. Unfortunately, Eq.(\ref{wdw}) makes no
reference to time, i.e., all the quantities entering it are
defined on the 3--dimensional hypersurface $\Sigma_t$. This is one
of the most obvious manifestations of the problem of time in
general relativity. The situation  could not be worse! We have a
quantum theory in which the main dynamical equation can be solved
without considering the evolution in time.

Some researchers interpreted this result as an indication of the
necessity of a completely different ``{\em timeless}" approach to
quantum theory \cite{rovelli,rovelli1,rovelli2}. This approach is
still under construction and although, in principle, some
conceptual problems can be solved some other problems related to
``time ordering" and ``time arbitrariness"  appear which are, at
best, as difficult as the above described problems of time.

On the other hand, the most propagated interpretation of the
problem of time of the Wheeler--DeWitt equation (\ref{wdw}) is
that time must be reintroduced into the quantum theory by means of
an auxiliary physical entity whose values can be correlated with
the values of other physical entities. This correlation allows in
principle to analyze the evolution of physical quantities with
respect to the ``{\em auxiliary internal time}".  Since there is
no clear definition of the auxiliary internal time, one can only
use the imagination to choose a quantity as the time parameter.
For instance, if we have a physical quantity which classically
depends linearly on time, it could be a good candidate for an
auxiliary internal time. Although the linearity seems to be a
reasonable criterion, it is not a necessary condition. Examples of
this type of auxiliary internal time are the very well analyzed
minisuperspaces of quantum cosmology. In particular, one could
select the auxiliary internal time as one of the scale factors of
homogeneous cosmological models. The volume element which is a
combination of scale factors would be  also a good choice since in
most cases it evolves in cosmic time and reproduces the main
aspects of cosmological evolution. The volume element has also
been used recently in loop quantum cosmology \cite{boj,boj1}.
Certain low energy limits in string theory contain a tachionic
field which linearly evolves in time and, consequently, could be
used as auxiliary internal time for quantization \cite{sen}. We
will consider these examples with some more details in section
\ref{sec:probtime}.

Nevertheless, it is not clear at all if the procedure of fixing an
auxiliary internal time  can be performed in an exact manner and,
if it can be done, whether the results of choosing different
auxiliary times can be compared and are somehow related. Finally,
a most controversial point is whether such an auxiliary time can
be used to relate the usual concepts of spacetime.

In the last section we mentioned that the canonical quantization
procedure implies that the fields to be quantized are defined on a
background spacetime. In quantum field theory, the Minkowski
spacetime with its set of preferred inertial frames plays the role
of background spacetime. In general relativity there is  no place
for a background metric. In fact, the entries of  the metric are
the  physical entities we need to quantize. This rises a new
problem. If we success in quantizing the spacetime metric, we will
obtain quantum fluctuations of the metric which make impossible
the definition of spacelike, null or timelike intervals. But the
starting commutation relations require the existence of a
well--defined spacetime interval. For instance, the first
commutation relation of Eq.(\ref{comm}) is usually interpreted as
reflecting the fact that the points ${\bf x}$ and ${\bf x}^\prime$
are separated by a spacelike interval. However, there is no
background metric to define this causal structure. Moreover, if we
would choose an arbitrary background metric, the quantum
fluctuations of that metric could completely change the causal
character of the interval. So we are in a situation in which if we
want to solve the original problem, we must violate one the most
important postulates needed to find the solution. Obviously, this
is not a good situation to begin with.

\section{Canonical quantization in minisuperspace}
\label{sec:mini}

The first attempt at minisuperspace quantization is due to DeWitt
\cite{dw67}, although the concept of minisuperspace was introduced
by Misner \cite{misner} some years later. At that time Wheeler
\cite{wh} suggested the idea of superspace as the space of all
three--geometries as the arena in which the geometrodynamics
develops. A particular four--geometry being a trajectory in this
space. Later, Misner applied the Hamiltonian formulation of
general relativity to cosmological models, having in mind the
quantization of these cosmological models. He introduced the
concept of minisuperspace and minisuperspace quantization or
quantum cosmology to describe the evolution of cosmological
spacetimes as trajectories in the finite dimensional sector of the
superspace related to the finite number of parameters, needed to
describe the $t=const.$ slices of the models and the quantum
version of such models, respectively.

In the early 70's the minisuperspace models and their quantum
version were extensively studied, however, the interest in them
decreased at the middle of this decade till Hartle and Hawking
\cite{hhw} revived the field in the early 80's emphasizing the
path--integral approaches. This started a lively resurgence of
interest in minisuperspace quantization.

In 1987 Mac\'{\i}as, Obreg\'on, and Ryan \cite{mor} introduced the
supersymmetric quantum cosmology approach by applying ($N=1$)
supergravity to quantum minisuperspaces in order to obtain the
square root of the Wheeler--DeWitt equation, which governs the
evolution of the quantum cosmological models in the standard
approach. In 1988 D'Eath and Hughes \cite{death1} constructed a
locally supersymmetric 1--dimensional model for quantum cosmology,
based on a particular case of the Friedmann--Robertson--Walker
spacetime (see also \cite{death2}). Later on, these results where
generalized to include Bianchi cosmological models, supersymmetric
matter,  and cosmological constant \cite{death3,death4,death5}.

In 1994 Carrol, Freedman, Ort\'{\i}z, and Page \cite{cfop}, showed
that there is no--physical states in $N=1$ supergravity, unless
there exist an infinite set of gravitino modes. In 1998
Mac\'{\i}as, Mielke, and Socorro \cite{mamilo} showed that there
are no--physical states in supersymmetric quantum cosmology.

As stated in \cite{kura89}, one of the greatest difficulties with
quantum cosmology has always been the seductive character of its
results. It is obvious that taking the metric of a cosmological
model, which is truncated by an enormous degree of imposed
symmetry and simply plugging it into a quantization procedure
cannot give an answer that can be in any way interpreted as a
quantum gravity solution. What people do is to assume that one can
represent the metric as a series expansion in space dependent
modes, the cosmological model being the homogeneous mode, and that
in some sense one can ignore the dependence of the state function
on all inhomogeneous modes. This artificial freezing of modes
before quantization represents an obvious violation of the
uncertainty principle and cannot lead to an exact solution of the
full theory. However, the results of applying this untenable
quantization procedure have always seemed to predict such
reasonable and internally consistent behavior of the universe that
it has been difficult to believe that they have no physical
content.

The minisuperspace is often known as the homogeneous cosmology
sector, as mentioned above, infinitely many degrees of freedom are
artificially frozen by symmetries. This reduction is so drastic
that only an unphysical finite number of degrees of freedom is
left. The requirement of homogeneity limits the allowed
hypersurfaces to the leaves of a privileged foliation, which is
labeled by a single ``time" variable. One can parametrize such
hypersurfaces of homogeneity by the standard Euler angles
coordinates and characterize the spatial metric uniquely by three
real parameters, $\Omega$, $ \beta_{\pm}$. The $\Omega$ is related
to the volume of the hypersurface $\Sigma$ as follows:
\begin{equation}
\Omega = \ln \int_\Sigma d^3x \vert q(x)\vert^{1/2}\, .
\end{equation}
The $\beta$ parameters describe the anisotropy of the hypersurface
$\Sigma$. Due to the symmetry of the model, the super--momentum
constraints are identically satisfied, while the
super--Hamiltonian constraint reduces to:
\begin{equation}
H _\bot  = - p_\Omega^2 + p_+^2 + p_-^2+
\exp({-\Omega})\left[V(\beta_+,\beta_-) - 1\right] =0\, .
\end{equation}
The potential $\left[V(\beta_+,\beta_-) - 1\right]$ is a
combination of exponential terms, it vanishes at the origin and it
is positive outside of it \cite{ryan}. The parameter $\Omega$ is
usually considered as a kind of ``auxiliary internal time" (see
section \ref{sec:probtime}). A systematic analysis of the global
time problem for homogeneous cosmological models seems to lead,
quite generally, to the lack of a global time function. Even the
volume time $\Omega$ is not globally permitted, for instance in
oscillating models, since the universe would attend a given value
$\Omega < \Omega_{max}$ at least twice, once when expanding, and
once when recontracting.

Additionally, the Wheeler--DeWitt equation based on one particular
choice of time variable, like $\Omega$ in this case, may give a
different quantum theory than the same equation based on another
choice of the time variable. This is what Kucha\v{r} called {\em
the multiple choice problem} \cite{Kuchar}.

It is dangerous to draw conclusions from minisuperspace models to
full quantum gravity. Minisuperspace spacetimes possess a
privileged foliation by leaves of homogeneity which does not exist
in a generic spacetime. Kucha\v{r} and Ryan \cite{kura89} showed
that even in the simple case of a microsuperpace (a reduced
minisuperspace) the result of canonical quantization is not
related to the quantization of the seed minisuperspace. One should
try to avoid common practice, which consists of solving a time
problem for a model way down in the hierarchy, and jumping to the
conclusion that the time problems of quantum gravity are removed
by the same treatment.

\section{Canonical quantization in midisuperspace}
\label{sec:midi}

The simplest generalization of the homogeneous models are the
Gowdy cosmological models, since they possesses two Killing
vectors and therefore two ignorable coordinates, reducing the
problem to time (as in standard quantum cosmology) and one spacial
coordinate, which completely eliminates homogeneity and leads to a
system with an infinite number of degrees of freedom, i.e. a true
field theory on a midisuperspace.  Gowdy cosmologies are widely
studied midisuperspace models.

Moreover, the canonical quantization of $N=1$ supergravity  in the
case of a midisuperspace described by Gowdy $T^3$ cosmological
models has been already studied in \cite{gowsg}. The quantum
constraints were analyzed and the wave function of the universe
was derived explicitly. Unlike the minisuperspace case, it was
shown that physical states in midisuperspace models do exist. The
analysis of the wave function of the universe leads to the
conclusion that the classical curvature singularity present in the
evolution of Gowdy models is removed at the quantum level due to
the presence of the Rarita--Schwinger field. Since this
supegravity midisuperspace model shares the same problem as other
midisuperspace models, which consists in the lacking  of a
well--defined time parameter, in this work a classical solution
was used to drive the evolution in time.

The midisuperspace models provide a canonical description of
Einstein spacetimes with a group of isometries. Symmetries remove
infinitely many degrees of freedom of the gravitational field, but
there remain still infinitely many degrees of freedom. In spite of
this simplification, the midisuperspace constraints of general
relativity are still complicated functionals of the canonical
variables.

The study of midisuperspace models and covariant field systems,
like string models, indicates that if there exists an auxiliary
internal time which converts the old constraints of general
relativity into a Schr\"odinger equation form, such a time
variable is non--local functional of the geometric variables.

The Gowdy $T^3$ cosmological models have been analyzed in the
context of non--perturbative canonical quantization of gravity
\cite{pierri,mena,jer}. The arbitrariness in the selection of a
time parameter is a problem that immediately appears in the
process of quantization. For a specific choice of time it was
shown that there does not exist an unitary operator that could be
used to generate the corresponding quantum evolution. Therefore,
even in the case of midisuperspace models there is no natural time
parameter.

\section{The problem of time}
\label{sec:probtime}

Quite a lot of different proposals have been made over the years
on how to interpret time in quantum gravity, i.e., the time
paradox. Kucha\v{r} \cite{Kuchar} classified them in three basic
approaches. It should be stressed that the boundaries of these
interpretations are not clearly defined:

\begin{enumerate}

\item {\em Internal Time.} Time is hidden among the canonical
variables and it should be identified prior to quantization. The
basic equation upon the interpretation is based in a Schr\"odinger
equation, not a Wheeler--DeWitt one. Nevertheless, this
interpretation is susceptible to the multiple choice problem,
i.e., the Schr\"odinger equation based on different time variables
may give different quantizations.
\begin{enumerate}

\item {\em Matter clocks and reference fluids.} The standard of
time is provided by a matter system coupled to geometry, instead
by the geometry itself. The intrinsic geometry and extrinsic
curvature of a spacelike hypersurface enter into the constraints
of general relativity in a very complicated way. Nothing in the
structure of the mentioned constraints tell us how to distinguish
the true dynamical degrees of freedom from the quantities which
determine the hypersurface. The founding fathers of general
relativity suggested a conceptual devise which leads exactly to
that, i.e., the reference fluid. The particles of the fluid
identify space points and clocks carried by them identify instants
of time. This fixes the reference frame and the time foliation. In
this frame and on the foliation, the metric rather than the
geometry becomes measurable. The concept of reference fluid goes
back to Einstein \cite{eins}, and to Hilbert \cite{hilbert17} who
formalized the idea that the coordinate system should be realized
by a realistic fluid carrying clocks which keep a causal time.
They imposed a set of inequalities ensuring that the worldlines of
the reference fluid be timelike and the leaves of the time
foliation be spacelike.

The reference fluid is traditionally considered as a tenuous
material system whose back reaction on the geometry can be
neglected. There is just matter but not enough to disturb the
geometry. Instead of deriving the motion of the fluid from its
action, one encodes it in coordinate conditions. Those are
statements on the metric which holds in the coordinate system of
the fluid and are violated in any other coordinate system. Such
standpoint makes difficult to consider the reference fluid as a
physical object which in quantum gravity could assume the role of
an apparatus for identifying spacetime events.

In order to turn the reference fluid into a physical system, it is
possible to picture the fluid as a realistic material medium and
devise a Lagrangian which describes its properties. By adding this
Lagrangian to the Einstein--Hilbert Lagrangian, the fluid is
coupled to gravity. Other possibility is to impose the coordinate
condition before variation by adjoining them to the action via
Lagrange multipliers. The additional terms in the action are
parameterized and interpreted as matter source.

\item {\em Cosmological time.} In one special case, the reference
fluid associated with a coordinate condition allows a geometrical
interpretation. This is the unimodular coordinate condition, i.e.,
$\vert g_{\mu\nu} \vert^{1/2} = 1$, fixing the spacetime volume
element. These unimodular coordinates were proposed by Einstein
\cite{eins1}. By imposing the unimodular condition before rather
than after variation, a law of gravitation with unspecified
cosmological constant is obtained \cite{unruh}, reducing the
reference fluid to a cosmological term. The cosmological constant
appears as a canonical conjugate momentum to a time coordinate,
i.e., the cosmological time.

The path--integral version of this approach has been used by
Sorkin \cite{sorkin1,sorkin2} to show that in a simple model of
unimodular quantum cosmology the wave function remains regular as
the radius of the universe approaches the classical singularity,
but its evolution is non--unitary.

Moreover, it has been shown by Heneaux and Teilteboim
\cite{hennteit89} that the increment of the cosmological time
equals the four--volume enclosed between the initial and the final
hypersurfaces.

Unruh and Wald \cite{unruhwald} suggested that any reasonable
quantum theory should contain a parameter, called Heraclitian
time, whose role is to set the conditions for measuring quantum
variables and to provide the temporal order of such measurements.
The problem with this suggestion is that the cosmological time is
not in any obvious way related with the standard concept of time
in relativity theory. The basic canonical variables, the metric
$q_{ij}$ and its conjugate momentum $\pi^{ij}$ are always imposed
to be measured on a single spacelike hypersurface rather than at a
single cosmological time. In order to be able to introduce a
particular hypersurface, one needs to specify functions of three
coordinates, instead of a single real parameter, i.e., the
absolute time of Newtonian mechanics. Consequently, it remains the
question in what sense the cosmological time sets the conditions
of quantum measurements.

The cosmological time does not fix the conditions for a
measurement uniquely, since it it cannot differentiate between the
infinitely many possible hypersurfaces of the equivalence class,
in order to know in which one the geometrical variables are to be
measured. In other words, the hypersurfaces parametrized with
different values of the cosmological time are allowed to intersect
and cannot be causal ordered as the Heraclitian time requires.
Therefore, cosmological time (Heneaux--Teilteboim volume) is not a
functional time. Relativity time is a collection of all spacelike
hypersurfaces and no single parameter is able to label uniquely so
many events.

\item {\em Time and tachyons.} The specific form of the low energy
action of the tachyon dynamics reads \cite{sen}:
\begin{equation}
S= - \int d^{p+1} x V(T) \sqrt{1 + \eta^{\mu\nu} \partial_\mu T
\partial_\nu T}\label{tachion}\, ,
\end{equation}
where $p=9$ for strings type IIA or IIB, and $p=25$ for bosonic
strings, $\eta^{\mu\nu} = {\rm diag(-1,1,1, ... ,1)}$, and $V(T)$
is the potential of the tachyon $T$.

Sen \cite{sen} proposed that, at the classical level, solutions of
the equations of motion of the field theory described by
(\ref{tachion}), at ``{\em late time}" are in one to one
correspondence with configuration of non--rotating,
non--interacting dust. At ``{\em late time}" the classical vacuum
solutions of the equations of motion approach $T=x^0=$ time
coordinate, making $T$ a candidate for describing time at the
classical level. On the other hand, at ``{\em late time}" the
quantum theory of the tachyon $T$ coupled to gravity leads to a
Wheeler--DeWitt equation independent of $T$, whereas for ``{\em
early time}" or ``{\em finite time}" the resulting Wheeler--DeWitt
equation has a non--trivial dependence on $T$ in the considered
region.

Nevertheless, it is well known that the classical tachyon
dynamics, when quantized coupled to gravity, may not describe
correctly the physics arising from the quantum string theory.
Additionally, since the tachyon is identified with a configuration
of non--rotating and incoherent dust, its role as time variable
shares all the diseases, mentioned above, of reference fluids.
Therefore, even in string theory the time paradox remains
unsolved.

\end{enumerate}

\item {\em Wheeler--DeWitt framework.} The constraints are imposed
in the metric representation leading to a Wheeler--DeWitt
equation. The dynamical interpretation asserts that the solutions
would be insensitive to the time identification among the metric
functions. This interpretation has to deal with the fact that the
Wheeler--DeWitt equation presents familiar problems when one tries
to turn the space of its solutions into a Hilbert space. Hence,
the statistical interpretation of the theory is based on the inner
product. Moreover, if there is a Killing vector, no energy
operator commutes with the general relativity constraint $H_\bot$,
and the construction of the Hilbert space fails. Even if there
exists a timelike Killing vector, the positivity of the inner
product requires that the potential in the super--Hamiltonian is
non--negative.

The semiclassical interpretation hides the problem of time behind
an approximation procedure. It claims that the Wheeler--DeWitt
equation for a semiclassical state approximately reduces to the
Schr\"odinger equation, and the Klein--Gordon norm reduces to the
Schr\"odinger norm. Unfortunately, it achieves the positivity of
the norm at an unacceptable price of suspending the superposition
principle \cite{page,page1}. When the semiclassical interpretation
is applied to quantum gravity properly the problem of separating
the classical modes defining time from the quantum modes arises.
In other words, this means that quantum gravity would have a
probabilistic interpretation only if it is classical.

\item {\em Quantum gravity without time.} This interpretation
claims that one does not need time to interpret quantum gravity
and quantum mechanics in general. Time may emerge in particular
situations, but even if it does not, quantum states still allow a
probabilistic interpretation \cite{rovelli,rovelli1,rovelli2}.

Its difficulty stands on the fact how to explain quantum dynamics
in terms of constants of motion. The existing proposals are
ambiguous, since the replacement of the classical global time
parameter by an operator is ambiguous and its consequences lead to
the multiple choice problem and to the problem of how to construct
a Hilbert space \cite{Kuchar}.

As it is well known, in canonical formalism, gauge transformations
are generated by constraints linear in the momenta, and they move
a point in the phase space along, to what is usually called, an
orbit of the gauge group. Moreover, two points on the same orbit
are physically indistinguishable and represent two equivalent
descriptions of same physical state. An observable should not
depend on description of the chosen state, the state must be the
same along the given orbit, i.e., its Poisson bracket with all the
constraints must vanish.

On one hand, all the physical content of general relativity is
contained in the constraints and the observables are those
dynamical variables that have vanishing Poisson brackets with all
constraints. In particular, due to the fact that the
diffeormorphims constraint generates a gauge, i.e., the group of
spatial coordinates diffeomorphisms. Therefore, any observable in
general relativity must be invariant under diffeomorphisms.

On the other hand, the super--Hamiltonian constraint generates the
dynamical change of the geometrical dynamical variables from one
hypersurface to another, i.e., any dynamical variable which
commutes with the super--Hamiltonian must be the same on every
hypersurface and it must be constant of motion. Nevertheless, in
order to be able to maintain that the quantum observables are
those which commute with all the constraints of general relativity
seems to imply that our quantum universe can never change. The
transformations generated by the super--Hamiltonian should not be
interpreted as gauge transformations. Two points on the same orbit
of the super--Hamiltonian transformations are two events in the
dynamical evolution of the system which are distinguishable
instead of been two descriptions of the same physical state.

{\em Second quantization.} There exists a belief that the second
quantization solves the problem of time in quantum theory of a
relativistic particle. The second quantization approach to quantum
field theory is based on the construction of a Fock space, i.e.,
one takes a one--particle Hilbert space $F_{(1)}$. From the direct
product of the one--particle states, the states which span the
N--particle sector $F_{(N)}$ are constructed. The Fock space $F$
is then the direct sum of all such sectors, i.e., $F = F_{(0)}
\oplus F_{(1)} \oplus F_{(2)} \oplus \cdots$, where $F_{(0)}$ is
spanned by the vacuum state. It is clear that the Fock space $F$
can be defined only if the one--particle state $F_{(1)}$ is a
Hilbert space. This brings us to the Hilbert space problem for a
relativistic particle. The absence of a privileged one--particle
Hilbert space structure is the source of ambiguities in
constructing a unique quantum field theory on a dynamical
background \cite{Kuchar}.

A closer look to the second quantization approach reveals that it
does not really solve the problem of time evolution and its
formalism resists an operational interpretation, like the problems
presented by the indefinite inner product of the Klein--Gordon
interpretation, which are faced by suggesting that the solutions
of the Wheeler--DeWitt equation are to be turned to operator. This
is analogous to subjecting the relativistic particle, whose state
is described by the Klein--Gordon equation, to second
quantization.

In full, the second quantization merely shifts the problem of time
to a different level without really solving it.

\end{enumerate}

\section{Conclusions}
\label{sec:con} Since the concepts of time in quantum mechanics
and general relativity are drastically different from each other,
generalizations of the usual quantum theory are required to deal
with quantum spacetime and quantum cosmology \cite{hartle}. That
is due to the fact that the usual framework for quantum theory
assumes a fixed background spacetime geometry. Physical states are
defined on spacelike hypersurfaces in this geometry and evolve
unitarily in between such hypersurfaces in the absence of
measurements and by state vector reduction on them when a
measurement occurs. The inner product is defined by integrals over
fields on a spacelike hypersurface. Nevertheless, at the quantum
realm, spacetime geometry is not fixed, but a dynamical variable
fluctuating and without definite value. It is not possible to
determine whether two given nearby points on a spacetime manifold
are spacelike separated or not. Instead, the amplitudes for
predictions are sums over different metrics on the manifold.
Additionally, points separated by a spacelike intervals in one
metric could be timelike separated in another metric, that
contributes just as significantly to the sum. Moreover, quantum
theory does not provide a natural time parameter and the quantum
constraints of general relativity do not contain any time
parameter. For this reason, standard quantum mechanics needs to be
generalized to accommodate quantum spacetime.

On the other hand, the application of quantum mechanics to quantum
cosmology also requires another kind of generalization of the
standard formulation. Standard quantum mechanics predicts the
outcome of measurements carried out on a system by another system
outside it. However, in cosmology there is no outside. Therefore,
quantum cosmology requires a quantum mechanics for closed systems,
i.e., a generalization of the standard quantum theory.

All the attempts to implement the canonical quantization procedure
to quantize systems in which time is not Newtonian do not provide
a reasonable description of the corresponding quantum system. The
quantization of general relativity has been an open problem for
more than 70 years and the leading present approaches, string
theory and loop quantum gravity, are far from providing an
ultimate solution, although many technical problems have been
attacked and partially solved in the past 20 years. Nevertheless,
it seems that the main conceptual problems, especially the one
related to time, are still not well understood. In our opinion, it
is not possible to reconciliate and integrate into a common scheme
the absolute and non--dynamical character of Newtonian time of
canonical quantization with the relativistic and dynamical
character of time in general relativity. What is needed is a
radical change of perspective either in general relativity or in
quantum mechanics. That is to say, we need either a theory of
gravity with an non--dynamical Newtonian time or a quantum theory
with a dynamical time in its construction. We believe that what
requires radical changes is the canonical quantization procedure
in such a way that the concept of time enters it in a  more
flexible manner. The issue of time remains as an open problem.



\subsection*{Acknowledgment}

One of us (A.M.) thanks Claus L\"ammerzahl and Bertfried Fauser
for enlightening discussions in Bremen and Blaubeuren,
respectively. This work was supported by CONACyT grants 48601--F,
and 48404--F.
\end{document}